\documentclass[11pt,a4paper]{article}

\usepackage{amsmath}
\usepackage{amsthm,amssymb}

\usepackage{graphicx} 

\newcommand{\beq}{\begin{equation}}
\newcommand{\eeq}{\end{equation}}

\numberwithin{equation}{section}

\begin{document}
\title{Exact Solution of the Discrete (1+1)--dimensional RSOS Model with Field and Surface Interactions}
\author{A L Owczarek$^1$ and T Prellberg$^2$\\
  \footnotesize
  \begin{minipage}{13cm}
    $^1$ Department of Mathematics and Statistics,\\
    The University of Melbourne, Parkville, Vic 3010, Australia.\\
    \texttt{owczarek@unimelb.edu.au}\\[1ex] 
$^2$ School of Mathematical Sciences\\
Queen Mary University of London\\
Mile End Road, London E1 4NS, UK\\
\texttt{t.prellberg@qmul.ac.uk}
\end{minipage}
}

\maketitle  

\begin{abstract}

We present the solution of a linear Restricted Solid--on--Solid (RSOS) model in
a field. Aside from the origins of this model in the context of describing the 
phase boundary in a magnet, interest also comes from more recent work on the 
steady state of non-equilibrium models of molecular motors.

While similar to a previously solved (non-restricted) SOS model in its physical 
behaviour, mathematically the solution is more complex. Involving basic
hypergeometric functions ${}_3\phi_2$, it introduces a new form of solution to
the lexicon of directed lattice path generating functions. 

\end{abstract}

\section{Introduction}

The SOS model arose from the consideration of the boundary between oppositely 
magnetised phases in the Ising model \cite{te52} at low temperatures and is now considered
to be useful for describing the salient features of a wide variety of interfacial
phenomena \cite{di88,svprab88,fi84,ab83,abow90}. The configurations involved 
in the linear (1+1)--dimensional case, modelling the interface in a two-dimensional 
magnet, have also been used to model the backbone of a polymer in solution. 
The critical phenomena associated with this model describe wetting transitions of
the interface with a wall \cite{di88}.
For the SOS model the phase diagram contains a wetting transition at
finite temperature $T_w$ for zero field and complete wetting occurs taking
the limit $H\to0$ for $T\geq T_w$ \cite{owczarek93a}. 

The linear SOS model with magnetic field and wall interaction was solved in \cite{owczarek93a}.
The restricted SOS model is a variant of the SOS model where the interface takes
on a restricted subset of configurations as opposed to the full SOS model. Effectively
it suppresses very large local fluctuations of the interface. This model
has been considered with wall interaction, but not yet with magnetic field. This is partially because,
with wall interaction only, the SOS and RSOS models are in the same universality class, as
demonstrated by their exact solutions \cite{prsv89,chwe81,ch81,bu81,hiva81,kr81}. Recently, 
it has been suggested that the RSOS model in a field describes the steady state of a non-equilibrium model of 
a molecular motor \cite{marko}. This observation has motivated us to consider the RSOS model in a field.
In doing so, we have discovered a novel form of generating function for a directed lattice path problem,
and a new method of solution for such problems.

The RSOS model we analyse can be described as follows. Consider a two-dimensional square lattice 
in a half plane. For each column $i$ of the
surface a segment of the interface is placed on the horizontal link at height $r_i\geq0$ and
successive segments are joined by vertical segments to form a partially directed interface
with no overhangs. The configurations are given the energy
\beq
-\beta E=-K\sum_i|r_i-r_{i-1}|-H\sum_ir_i+b\sum_i\delta_{r_i,0} \; .
\eeq
There are two basic variants of this model which have been discussed in
the literature. If there are no restrictions placed on the differences of
successive heights $r_i$, the model is called the (unrestricted) SOS model,
analysed in \cite{owczarek93a}.

On the other hand, constraining the height differences to be bounded by 
\beq
|\Delta r_i|=|r_i-r_{i-1}|\leq1
\eeq
gives the {\em restricted} SOS (RSOS) model. This has previously only been considered in 
the case of zero field $H$ \cite{prsv89,chwe81}, for several types of external potential 
\cite{prsv88}. 
Both variants have been
considered, utilising a different thermodynamic ensemble, as models for
polymers in solution, since the finite configurations are partially directed self-avoiding walks
\cite{prsv89,fo89}.  In \cite {fo91} a RSOS model with $H=0$
but a rigidity term dependent on $|\Delta r_i-\Delta r_{i-1}|$ has
been considered as a model of semi-flexible polymers such as DNA.

\section{The RSOS generating function}

\begin{figure}[ht]
\begin{center}
\includegraphics[width=0.9\textwidth]{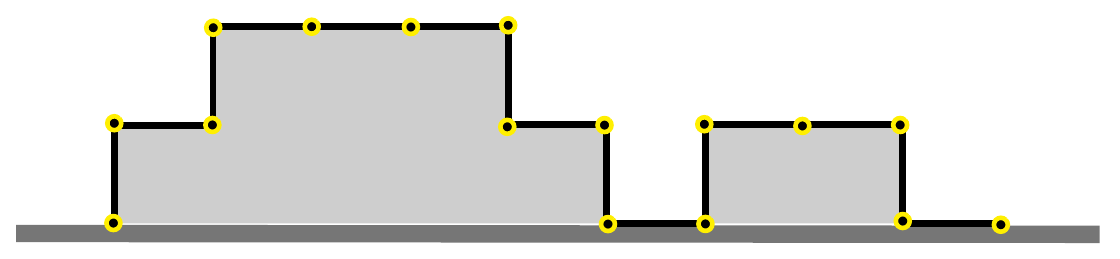}
\caption{A typical RSOS configuration beginning on the surface and finishing on the surface with
a horizontal step: each horizontal step is assigned a weight $x$, each vertical step a weight $y$,
each unit of area a weight $q$, and each step that touches the surface an additional weight $\kappa$.
The width of the configuration shown here is $N=9$, and the heights are $r_0=0$, $r_1=1$, $r_2=r_3=r_4=2$, 
$r_5=1$, $r_6=0$, $r_7=r_8=1$, and $r_9=0$. The weight of this configuration equals $x^9y^6q^{10}\kappa^2$.}
\label{figure1}
\end{center}
\end{figure}

We discuss the RSOS model in terms of lattice paths.
An RSOS path is a partially directed self-avoiding path with no steps into
the negative $x$-direction and no successive vertical steps. To be precise,
an RSOS path of width $N$ with heights $r_0$ to $r_N$ has horizontal steps at heights
$r_1,\ldots,r_N$, and vertical steps between heights $r_{i-1}$ and $r_i$ for $i=1,\ldots,N$,
but no horizontal step associated with $r_0$. This means that an RSOS path starts at height $r_0$ 
with either a horizontal step (if $r_1=r_0$) or vertical step (if $r_1\neq r_0$), but must 
end at height $r_N$ with a horizontal step. Figure \ref{figure1} shows an example.

The partition function for the RSOS paths of width $N$ with ends fixed at heights
$r_0\geq0$ and $r_N\geq0$, respectively, is given by
\beq
Z_1(r_0;r_1)=\begin{cases}\exp(-\beta E(r_0;r_1))\;,&|r_0-r_1|\leq 1\\ 0\;, &|r_0-r_1|>1\;,\end{cases}
\eeq
and
\beq
Z_N(r_0;r_N)=\sum_{\substack{r_1,\ldots,r_{N-1}\geq0\\|r_i-r_{i-1}|\leq1}}
\exp(-\beta E(r_0;r_1,\ldots,r_N))\;,
\qquad N=2,3,\ldots \; ,
\eeq
where 
\beq
-\beta E(r_0;r_1,\ldots,r_N)=
-K\sum_{i=1}^N|r_i-r_{i-1}|-H\sum_{i=1}^Nr_i+b\sum_{i=1}^N\delta_{r_i,0}
\; .
\eeq
Here, we shall consider paths with both ends attached to the surface, i.e. we
shall focus on the partition function 
\beq
Z_N=Z_N(0;0)\;.
\eeq
We define
\beq
y=\exp(-K)\;,\quad q=\exp(-H)\;,\quad\mbox{and}\quad\kappa=\exp(b)\;,
\eeq
so $y$ is a temperature--like, $q$ a magnetic field--like and $\kappa$ a
binding energy--like variable,
and write
\beq
Z_N=Z_N(y,q,\kappa)\;.
\eeq
The free energy is then 
\beq
- \beta f(y,q,\kappa)=\lim_{N\to\infty}\frac1N\log Z_N(y,q,\kappa)\;.
\eeq
Define the generalised (grand canonical) partition function, or
simply generating function, as
\beq
\label{partition_function}
G(x,y,q,\kappa)=1+\sum_{N=1}^\infty x^NZ_N(y,q,\kappa)\;.
\eeq
Thus, the radius of convergence $x_c(y,q,\kappa)$ of $G(x,y,q,\kappa)$ with 
respect to the series expansion in $x$ can be identified as
$\exp(\beta f(y,q,\kappa))$, hence
\beq
 f(y,q,\kappa) = kT \log x_c(y,q,\kappa)\;.
\eeq

It is convenient to consider $G$ as a combinatorial generating function for 
RSOS paths, where $x$, $y$, $q$ and $\kappa$ are counting variables for 
appropriate properties of those paths. Interpreted in such a way, $x$ and $y$ 
are the weights of horizontal and vertical steps, respectively, $q$ 
is the weight for each unit of area enclosed by the RSOS path and the $x$-axis, 
and $\kappa$ is is an additional weight for each step that touches the surface.
For example, the weight of the configuration in Figure \ref{figure1} is $x^9y^6q^{10}\kappa^2$.

We find easily the first few terms of $G$ as a series expansion in $x$,
\beq
G(x,y,q,\kappa)=1+\kappa x+(\kappa^2+\kappa y^2q)x^2+\ldots\;,
\eeq
where the constant term corresponds to a zero-step path starting and 
ending at height zero with weight one.

\section{Functional equation and exact solution}

\begin{figure}[ht]
\begin{center}
\includegraphics[width=\textwidth]{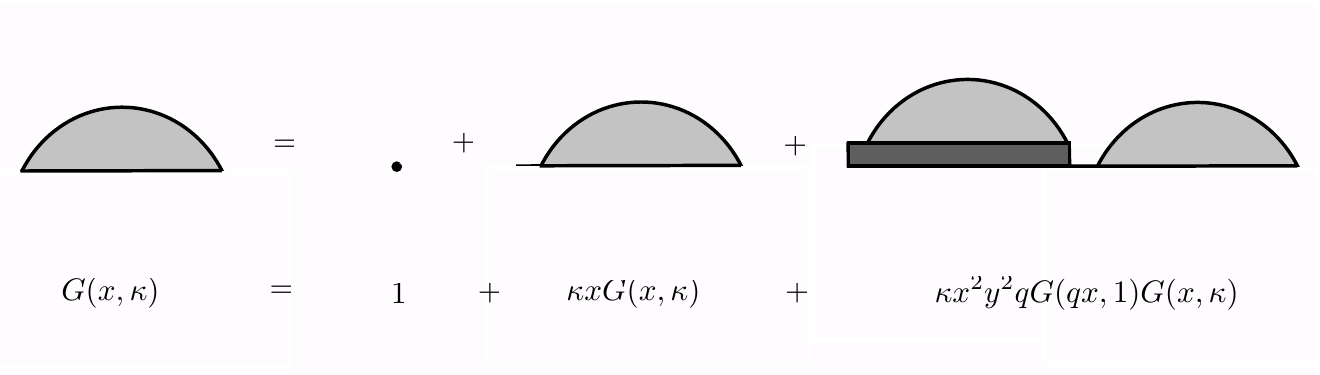}
\caption{The diagrammatic form of the functional equations for RSOS paths, indicating the
combinatorial decomposition of RSOS paths.}
\label{figure2}
\end{center}
\end{figure}

The key to the solution is a combinatorial decomposition of RSOS paths which leads to a
functional equation for the generating function $G$. This decomposition is done with
respect to the left-most horizontal step touching the surface at height zero, and is
shown diagrammatically in Figure \ref{figure2}.  

We distinguish three cases:
\begin{itemize}
\item[(a)] The RSOS path has zero width, and there is no horizontal step at height zero. The
contribution to the generating function is $1$.
\item[(b)] The RSOS path starts with a horizontal step, which therefore is at height zero. The rest of
this path is again a RSOS path. The contribution to the generating function is 
$\kappa xG(x,y,q,\kappa)$.
\item[(c)] The RSOS path starts with a vertical step. Then there will be a left-most horizontal step 
at height zero, and removing this step cuts the path into two pieces. The left path starts with a vertical 
and horizontal step, followed by an RSOS path starting and ending at height one and not touching the
surface, followed by a vertical step to height zero. The right path is again a RSOS path. The
contribution to the generating function is $yxqG(qx,y,q,1)y\kappa xG(x,y,q,\kappa)$.
\end{itemize}
Put together, this decomposition leads to a functional equation for the generating function 
\beq
\label{functeq}
G(x,y,q,\kappa)=1+\kappa xG(x,y,q,\kappa)+\kappa x^2y^2qG(qx,y,q,1)G(x,y,q,\kappa)\;.
\eeq
For $q=1$, the solution is a simple algebraic function
\beq
\label{qisone}
G(x,y,1,\kappa)=\left(1-\kappa x-\frac\kappa2\left(1-x-\sqrt{(1-x)^2-4x^2y^2}\right)\right)^{-1}\;,
\eeq
and for general values of $q$, a formal iteration of (\ref{functeq}) leads to
a continued fraction expansion
\beq
\label{confrac}
G(x,y,q,\kappa)=
\cfrac{1}{1-\kappa x-
\cfrac{\kappa qx^2y^2}{1-qx-
\cfrac{q^3x^2y^2}{1-q^2x-
\cfrac{q^5x^2y^2}{1-q^3x-
\cfrac{q^7x^2y^2}{1-q^4x-
\ldots}}}}}\;.
\eeq
However, there is a non-trivial method to solve the functional equation for $G$ in terms of power series. 
Our main result is an expression for $G$ involving $q$-series. In the next section, we shall
derive the following expression:
\beq
\label{mainresult}
G(x,y,q,1)=\frac{
\alpha\,(\lambda qx;q)_\infty\phi\left[\lambda qx,q\frac{\lambda}{1-\lambda};q,q\right]+((1-\lambda)qx;q)_\infty\phi\left[(1-\lambda)qx,q\frac{1-\lambda}\lambda;q,q\right]
}
{
\alpha\,(\lambda x;q)_\infty\phi\left[\lambda x,q\frac{\lambda}{1-\lambda};q,q\right]+((1-\lambda)x;q)_\infty\phi\left[(1-\lambda)x,q\frac{1-\lambda}\lambda;q,q\right]
}
\eeq
with
\beq
\alpha=\frac{
\phi\left[0,q\frac{1-\lambda}\lambda;q,q\right]-(1-\lambda)\phi\left[0,q\frac{1-\lambda}\lambda;q,q^2\right]
}{
\lambda\phi\left[0,q\frac\lambda{1-\lambda};q,q^2\right]-\phi\left[0,q\frac\lambda{1-\lambda};q,q\right]
}
\eeq
and 
\beq
y^2=\lambda(1-\lambda)\;.
\eeq
Here $\phi$ is given in terms of the basic hypergeometric function ${}_3\phi_2$ as
\beq
\phi[s,t;q,z]={}_3\phi_2\left(\genfrac{}{}{0pt}{}{0,0,0}{s,t};q,z\right)=\sum_{n=0}^\infty\frac{z^n}{(s;q)_n(t;q)_n(q;q)_n}
\eeq
and 
\beq
(t;q)_n=\prod_{k=0}^{n-1}(1-tq^k)
\eeq
is the standard $q$-product. From Equation (\ref{functeq}) we have the full solution as
\beq
G(x,y,q,\kappa)=\frac1{1-\kappa x-\kappa x^2y^2qG(qx,y,q,1)}\;.
\eeq

\section{Solving the functional equation}

Using a linearisation Ansatz \cite{pb95}, standard for a $q$-deformed algebraic equation such as 
Equation (\ref{functeq}), we substitute
\beq
\label{linearization}
G(x,y,q,1)=\frac{H(qx,y,q)}{H(x,y,q)}
\eeq
into Equation (\ref{functeq}) with $\kappa=1$, and find that $H(x,y,q)$ must satisfy the {\em linear} $q$-functional
equation
\beq
\label{Heqn}
H(x,y,q)+(x-1)H(qx,y,q)+qx^2y^2H(q^2x,y,q)=0\;.
\eeq
We then try to solve this linear functional equation using a series in $x$,
\beq
\label{Hansatz}
H(x,y,q)=\sum_{n=0}^\infty c_n(y,q)x^n\;.
\eeq
This unfortunately leads to a non-trivial three-term recurrence for the coefficients
\beq
\label{crecurrence}
(1-q^{n+2})c_{n+2}(y,q)+q^{n+1}c_{n+1}(y,q)+y^2q^{2n+1}c_n(y,q)=0
\eeq
for $n\geq0$ with initial condition $c_0(y,q)+(1-q)c_1(y,q)=0$.
This is different from the two-term recurrences obtained when considering the models 
in \cite{pb95} which can be solved by direct iteration. 

Inspired by the structure of basic hypergeometric functions, which we know form the
basis for a solution in the SOS model \cite{owczarek93a}, we transform the coefficients
as
\beq
\label{cansatz}
c_n(y,q)=\frac{(-1)^nq^{n(n-1)/2}}{(q;q)_n}d_n(y,q)\;.
\eeq
This leads to a recurrence
\beq
\label{recurrence}
d_{n+2}(y,q)-d_{n+1}(y,q)+y^2d_n(y,q)=y^2q^{n+1}d_n(y,q)
\eeq
for $n\geq0$ with initial condition $d_1(y,q)=d_0(y,q)$. While this is still a three-term
recurrence, only one of the terms has a non-constant coefficient. 

The left-hand side of Equation (\ref{recurrence}) is a homogeneous difference equation with 
constant coefficients and characteristic polynomial
\beq
\label{charpoly}
P(\lambda)=\lambda^2-\lambda+y^2\;.
\eeq
If the right-hand side of Equation (\ref{recurrence}) were zero, the solution would be
given as $d_n=A\lambda_1^n+B\lambda_2^n$ where $\lambda_i$ are the roots of $P(\lambda)=0$.
To solve this recurrence, we use the Ansatz \cite{prsv89}
\beq
\label{dansatz}
d_n(y,q)=\lambda^n\sum_{m=0}^\infty e_m(y,q,\lambda)q^{nm}\;.
\eeq
Inserting this Ansatz into Equation (\ref{recurrence}) we find
\beq
P(\lambda)e_0(y,q,\lambda)+\sum_{m=1}^\infty q^{mn}\left[P(\lambda q^m)e_m(y,q,\lambda)-y^2qe_{m-1}(y,q,\lambda)\right]=0\;.
\eeq
Necessarily $P(\lambda)=0$, and normalising by letting $e_0(y,q,\lambda)=1$, we find by
iteration
\beq
e_m(y,q,\lambda)=\frac{q^m}{(\lambda^2q/y^2;q)_m(q;q)_m}\;.
\eeq
The full solution to the recurrence equation (\ref{recurrence}) is a linear combination of (\ref{dansatz})
over both values of $\lambda$ satisfying $P(\lambda)=0$. If $P(\lambda)=0$, then also $P(y^2/\lambda)=0$,
and we can write
\beq
\label{dsoln}
d_n(y,q)=A\lambda^n\sum_{m=0}^\infty e_m(y,q,\lambda)q^{nm}
+B(y^2/\lambda)^n\sum_{m=0}^\infty e_m(y,q,y^2/\lambda)q^{nm}\;,
\eeq
where $\lambda$ is an arbitrarily chosen solution of $\lambda(1-\lambda)=y^2$.
Using the initial condition $d_0(y,q)=d_1(y,q)$ we can solve for the ratio $\alpha=A/B$.
We can somewhat simplify our expressions by noting that the dependence between
$\lambda$ and $y$ implies that we can replace $y^2$ by $\lambda(1-\lambda)$.
Noticing that the functions involved are ${}_2\phi_1$ basic hypergeometric series, we have
\beq
\label{alpha}
\alpha=\frac{
{}_2\phi_1\left(\genfrac{}{}{0pt}{}{0,0}{q\frac{1-\lambda}\lambda};q,q\right)
-(1-\lambda)
{}_2\phi_1\left(\genfrac{}{}{0pt}{}{0,0}{q\frac{1-\lambda}\lambda};q,q^2\right)
}{
\lambda
{}_2\phi_1\left(\genfrac{}{}{0pt}{}{0,0}{q\frac\lambda{1-\lambda}};q,q^2\right)
-
{}_2\phi_1\left(\genfrac{}{}{0pt}{}{0,0}{q\frac\lambda{1-\lambda}};q,q\right)
}
\eeq
where
\beq
{}_2\phi_1\left(\genfrac{}{}{0pt}{}{0,0}{s};q,z\right)=\sum_{n=0}^\infty\frac{z^n}{(s;q)_n(q;q)_n}=
{}_3\phi_2\left(\genfrac{}{}{0pt}{}{0,0,0}{0,s};q,z\right)\;.
\eeq
Substituting the expression for $d_n$ given in (\ref{dsoln}) back into $H$ using Equations (\ref{cansatz}) 
and (\ref{Hansatz}), we find that
\begin{align}
H(x,y,q)=&\alpha\sum_{m,n=0}^\infty\frac{q^{n(n-1)/2}(-\lambda x)^nq^{nm+m}}
{(q\lambda/(1-\lambda);q)_m(q;q)_m(q;q)_n}\\
&+\sum_{m,n=0}^\infty\frac{q^{n(n-1)/2}(-(1-\lambda)x)^nq^{nm+m}}
{(q(1-\lambda)/\lambda;q)_m(q;q)_m(q;q)_n}\;.
\nonumber\end{align}
The summation over $n$ can be done explicitly using Euler's formula \cite{gara90}
\beq
(t;q)_\infty=\sum_{n=0}^\infty\frac{q^{n(n-1)/2}(-t)^n}{(q;q)_n}\;.
\eeq
We find that
\beq
H(x,y,q)=\alpha\sum_{m=0}^\infty\frac{(\lambda xq^m;q)_\infty q^m}
{(q\lambda/(1-\lambda);q)_m(q;q)_m}
+\sum_{m=0}^\infty\frac{((1-\lambda)xq^m;q)_\infty q^m}
{(q(1-\lambda)/\lambda;q)_m(q;q)_m}
\eeq
which after pulling out a $q$-product factor from each sum can be identified with
\begin{align}
\label{Hsoln}
H(x,y,q)=&\alpha\,(\lambda x;q)_\infty
{}_3\phi_2\left(\genfrac{}{}{0pt}{}{0,0,0}{\lambda x,q\lambda/(1-\lambda)};q,q\right)\\
&+((1-\lambda)x;q)_\infty
{}_3\phi_2\left(\genfrac{}{}{0pt}{}{0,0,0}{(1-\lambda)x,q(1-\lambda)/\lambda};q,q\right)\;.
\nonumber
\end{align}
Substituting into the linearisation Ansatz (\ref{linearization}) gives us the solution
written in Equation (\ref{mainresult}).

\section{Comparison of SOS and RSOS model solutions}

The unrestricted SOS model was solved in \cite{owczarek93a} by a different technique,
namely the Temperley method. For the sake of comparison it is worthwhile reproducing 
the solution of the SOS model via the functional equation technique presented above 
for the restricted SOS model.
This highlights a fundamental difference in the difficulty of solving the two models.

In analogy to the functional equation (\ref{functeq}), the SOS model generating function 
$S(x,y,q,\kappa)$
satisfies
\beq
S(x,y,q,\kappa)=1+\kappa \left(x(1-y^2)-\frac{y^2}q\right)S(x,y,q,\kappa)+\kappa\frac{y^2}qS(qx,y,q,1)S(x,y,q,\kappa)\;,
\eeq
where the variables have identical meaning.

Setting $\kappa=1$ and substituting the linearisation Ansatz $S(x,y,q,1)=T(qx,y,q)/T(x,y,q)$ analogous to
the one used above in Equation (\ref{linearization}) yields
\beq
\label{Teqn}
T(x,y,q)+\left(x(1-y^2)-1-\frac{y^2}q\right)T(qx,y,q)+\frac{y^2}qT(q^2x,y,q)=0\;.
\eeq
While superficially this may seem very similar to the linear functional equation (\ref{Heqn}) for $H$,
it is important to recognize that the method of solution is to substitute a series in the variable $x$ into
the functional equation. 
Equation (\ref{Teqn}) above contains only linear factors in $x$, while  
the functional equation (\ref{Heqn}) contains factors that are quadratic in $x$,
This has the effect that the resulting 
recurrence for the coefficents of the series expansion of $T$ in $x$ is a two-term recurrence, while we found
a three-term recurrence in the case of the RSOS model. The two-term recurrence leads to an immediate solution
by iteration, while the three-term recurrence (\ref{crecurrence}) requires the extra work detailed in the 
previous section. One readily finds
\beq
T(x,y,q)=\sum_{n=0}^\infty\frac{(-1)^nq^{n(n-1)/2}[x(1-y^2)]^n}{(y^2;q)_n(q;q)_n}
={}_1\phi_1\left(\genfrac{}{}{0pt}{}{0}{y^2};q,x(1-y^2)\right)\;,
\eeq
which leads to 
an expression equivalent to Equation (45) in \cite{owczarek93a}. The resulting expression for $T$
should be compared with the much more complicated expression for $H$ given in Equations (\ref{Hsoln}) and
(\ref{alpha}).

\section{Conclusion}

In this paper we have presented a solution to the linear RSOS model in a field.
We have expressed the solution in terms of basic hypergeometric functions at values
of their arguments which are not powers of the counting variable in a combinatorial
problem. We know of one other case where this occurs in a different manner \cite{br03}.

We note that there is a bijection between RSOS paths and Motzkin paths. Hence
it would be interesting to consider other lattice path problems with a similar
structure such as $k$-coloured Motzkin paths \cite{bdpp95}.

It will also be interesting to see if this solution will produce insights into
non-equilibrium models of molecular motors.

\section*{Acknowledgements}

We thank J F Marko for suggesting this problem to us.
Financial support from the Australian Research Council via its support
for the Centre of Excellence for Mathematics and Statistics of Complex
Systems is gratefully acknowledged by the authors. A L Owczarek thanks the
School of Mathematical Sciences, Queen Mary, University of London for
hospitality.


\begin{thebibliography}{10}

\bibitem{te52}
H. N. V. Temperley,
Proc. Camb. Phil. Soc. {\bf 48}, 638 (1952).

\bibitem{di88} 
S. Dietrich, in {\em Phase Transitions and Critical Phenomena},
Vol. 12, ed.\ by C. Domb and J. L. Lebowitz, (Academic Press, London, 1988).

\bibitem{svprab88}
N. M. \v Svraki\'c, V. Privman, and D. B. Abraham,
J. Stat.\ Phys.\ {\bf 53}, 1041 (1988).

\bibitem{fi84}
M. E. Fisher,
Boltzmann Lecture in J.\ Stat.\ Phys.\ {\bf 34}, 667 (1984).

\bibitem{ab83}
D. B. Abraham,
Phys.\ Rev.\ Lett.\ {\bf 50}, 291 (1983).

\bibitem{abow90}
D. B. Abraham and A. L. Owczarek,
Phys.\ Rev.\ Lett.\ {\bf 64}, 2595 (1990).

\bibitem{owczarek93a}
{A. L. Owczarek and T. Prellberg, 
J. Stat. Phys. {\bf 70} 1175 (1993)}

\bibitem{prsv89} 
V. Privman  and N. M. \v Svraki\'c,
{\em Lecture Notes in Physics} {\bf 338} (Springer--Verlag, Berlin,
1989).

\bibitem{chwe81}
S. T. Chui and J. D. Weeks, 
Phys.\ Rev.\ {\bf B23}, 2438 (1981).

\bibitem{ch81}
J. T. Chalker,
J. Phys.\ A {\bf 14}, 2431 (1981).

\bibitem{bu81}
T. W. Burkhardt,
J. Phys.\ A {\bf 14}, L63 (1981).

\bibitem{hiva81}
H. Hilhorst and J. M. Van Leeuwin,
Physica {\bf 107A}, 319 (1981).

\bibitem{kr81}
D. M. Kroll,
Z. Phys.\ B {\bf 41}, 345 (1981).
 
\bibitem{marko}
{J. F. Marko,
private communication.}

\bibitem{prsv88}
V. Privman  and N. M. \v Svraki\'c,
J. Stat.\ Phys.\ {\bf 51}, 1111 (1988).

\bibitem{fo89} 
G. Forgacs, V. Privman, and H. L.Frisch,
J.\ Chem.\ Phys.\ {\bf 90}, 3339 (1989).

\bibitem{fo91}
G. Forgacs,
J. Phys.\ A {\bf 24}, 1099 (1991).

\bibitem{pb95}
{T. Prellberg and R. Brak, 
J. Stat. Phys. {\bf 78}, 701 (1995).}

\bibitem{gara90}
G. Gasper and M. Rahman, {\it Basic Hypergeometric Series} (Cambridge
University Press, 1990).

\bibitem{br03}
M. Bousquet-M\'elou and A. Rechnitzer,
Advances in Applied Mathematics {\bf 31} 86 (2003).

\bibitem{bdpp95}
E. Barcucci, A. Del Lungo, E. Pergola and R. Pinzani, 
{\em Proc. of the First Annual International Conference on Computing and Combinatorics} p.\ 254
(Springer, Berlin, 1995)

\end{thebibliography}
\end{document}